\RequirePackage{fix-cm}
\documentclass[smallextended]{svjour3}       
\usepackage{graphicx}
\usepackage{dcolumn}
\usepackage{bm}
\usepackage{graphicx}
\usepackage{latexsym,color}
\usepackage{amssymb}
\begin{document}
\newcommand{\ti}[1]{\mbox{\tiny{#1}}}
\newcommand{\im}{\mathop{\mathrm{Im}}}
\def\be{\begin{equation}}
\def\ee{\end{equation}}
\def\bea{\begin{eqnarray}}
\def\eea{\end{eqnarray}}
\newcommand{\tb}[1]{\textbf{\texttt{#1}}}
\newcommand{\rtb}[1]{\textcolor[rgb]{1.00,0.00,0.00}{\tb{#1}}}
\newcommand{\il}{~}
\newcommand{\rc}{\rho_{\ti{C}}}
\newcommand{\dd}{\mathcal{D}}
\newcommand{\lie}{\mathcal{L}}
\newcommand{\Tem}{T^{\rm{em}}}
\newcommand{\g}[1]{\Gamma^{\phantom\ #1}}
\title{On the evolution equations for ideal magnetohydrodynamics in curved spacetime}
\author{Daniela Pugliese         \and
        Juan A. Valiente Kroon}

\institute{Daniela Pugliese \at
              School of Mathematical Sciences, Queen Mary University of London\\
Mile End Road, London E1 4NS, UK\\
              \email{dpugliese@maths.qmul.ac.uk}           
           \and
           Juan A. Valiente Kroon \at
              School of Mathematical Sciences, Queen Mary University of London\\
Mile End Road, London E1 4NS, UK\\
               \email{j.a.valiente-kroon@qmul.ac.uk}           
               }%

\date{Received: date / Accepted: date}

%
\maketitle

\begin{abstract}
We examine the problem of the construction of a first order symmetric
hyperbolic evolution system for the Einstein-Maxwell-Euler system. Our
analysis is based on a $1+3$ tetrad formalism which makes use of the
components of the Weyl tensor as one of the unknowns. In order to
ensure the symmetric hyperbolicity of the evolution equations implied
by the Bianchi identity, we introduce a tensor of rank 3 corresponding
to the covariant derivative of the Faraday tensor. Our analysis
includes the case of a perfect fluid with infinite conductivity (ideal
magnetohydrodynamics) as a particular subcase.
\end{abstract}

\section{Introduction}
The present article is concerned with the construction of a first
order system of quasilinear hyperbolic evolution equations describing
a charged ideal perfect fluid coupled to the Einstein field equations
---the \emph{Einstein-Maxwell-Euler system}. As the fluid is charged,
one needs to bring into consideration the Maxwell equations to
describe the electromagnetic field produced by the charged fluid
flow. This system contains as an important subcase that of the
so-called \emph{ideal magnetohydrodynamics (MHD)} ---which applies applies
to the study of plasma in situations where a fluid
subjected to significant magnetic fields.

\medskip
As it is well known, General Relativity admits initial value problem
formulation whereby one prescribes certain initial data on a
3-dimensional hypersurface, and one purports to reconstruct the
spacetime associated to this initial data ---a so-called \emph{Cauchy problem}. The formulation of an
initial value problem is a necessary starting point for the
construction of numerical solutions to, say, the
Einstein-Maxwell-Euler system. The importance of a initial value
formulation of the system under consideration is not restricted to
numerical considerations, it is also a natural starting point for a
wide variety of analytical studies of the qualitative properties of
the solutions to the equations. \emph{In this article we approach the
  construction of the evolution equations of the
  Einstein-Maxwell-Euler system from the point of view of mathematical
Relativity}. Hence, the amenability of our analysis to analytic
considerations takes precedence over numerical considerations of implementability.
 Examples of qualitative aspects of
the solutions to, say, the Einstein-Maxwell-Euler system
requiring an suitable initial value formulation are the discussion of
local and global existence problems and the analysis of the stability
of certain reference solutions. Our initial value formulation of MHD
will be used elsewhere to discuss the stability of certain
spherically symmetric configurations using the methods of
\cite{KreLor98}.

\medskip
In order to formulate an initial value problem for the
Einstein-Maxwell-Euler system, one has to break the covariance of the
equations by introducing coordinates and choosing a preferred timelike
direction in the spacetime. This choice of coordinates and timelike
direction is usually known as a \emph{gauge choice}. In matter models
which contain fluids, there is a natural timelike direction ---that
given by the fluid flowlines. The reasons behind the need of a gauge
choice to formulate an initial value problem are technical: the
machinery of the theory of partial differential equations does not
apply to tensorial objects. As a consequence, the discussion of
initial value problems in General Relativity is inherently gauge
dependent.

\medskip
The motivation behind the gauge choice procedure discussed in the previous
paragraph is to deduce, from our basic set of (tensorial) equations, a
closed\footnote{\emph{Closed} in this context means that there should
  be as many equations as unknowns.} subsystem of evolution equations which is in what is called
\emph{symmetric hyperbolic} form ---see e.g. \cite{FriRen00} for
precise definitions. This procedure is sometimes called
\emph{hyperbolic reduction}. Once a symmetric hyperbolic system of
equations has been obtained then the machinery of the theory of
hyperbolic equations applies and ensures that for suitable
data prescribed on an initial hypersurface, the evolution equations have a local solution (i.e. a solution
that exists at least for for a small interval of time) and that the
solutions thus obtained is unique. The same theory also ensures that
the solutions obtained from the Cauchy problem depend continuously on the values of
initial data. An initial value problem with the properties described
in the previous paragraph is said to be \emph{well-posed}. It is
important to point out that if a certain reduction procedure does not
lead an hyperbolic system, this does not imply that the resulting
problem is not well posed ---this this would require a much more
detailed analysis. Symmetric hyperbolicity is just a necessary
condition for well-posedness.

\medskip
The purpose of the present article is to show that the
Einstein-Maxwell-Euler system admits a reduction procedure leading to
a system of symmetric hyperbolic evolution equations. This system of
equations is of first order. As a consequence of our construction one
automatically obtains a local existence and uniqueness result for the
system under consideration.

\medskip
The hyperbolic reduction procedure described in the present article
borrows from the discussion of the evolution equations for the
Einstein-Euler system by H. Friedrich in
\cite{Fri98c} ---see also \cite{FriRen00,Fri96} and \cite{Reu98}. In this
reference a Lagrangian gauge was used to undertake the hyperbolic
reduction and to obtain the desired evolution equations. The term
\emph{Lagrangian} means, in this context, that one uses the flow lines
of the fluid to construct the preferred time direction on which the
hyperbolic reduction is based. As it will be seen in the main body of
the article, this
choice has important technical advantages. The central
equation in this discussion is the \emph{Bianchi identity}. It
provides evolution equations for the components of the Weyl
tensor. The addition of electromagnetic interactions to Friedrich's
system destroys, in principle, the symmetric hyperbolic nature of the
evolution equation as derivatives of the Faraday tensor enter into the
principal part of the Bianchi evolution equations. This difficulty can
be handled by the introduction of a new field unknown corresponding to
the derivative of the Faraday tensor for which suitable field and
evolution equations can be obtained ---this idea is borrowed from the
analysis of the conformal Einstein-Maxwell system of \cite{Fri91}. The
system of equations considered are similar in spirit to those frame
formalism equations  of \cite{EllEls98}.

\medskip
The evolution equations for the Einstein-Maxwell-Euler system were
first considered from the perspective from the Cauchy problem in
\cite{Cho65} ---see also \cite{Cho08}. This analysis makes use of a
\emph{traditional} formulation of the Einstein field equations in
which the gravitational field is described by the metric tensor, and
uses wave coordinates to obtain non-linear wave equations for the
metric tensor. In this approach, the resulting evolution system is of
mixed order. As a consequence, the so-called \emph{Leray theory} needs
to be used ---see \cite{Cho08}. As pointed out in that reference, a
first order formulation of the problem does not necessarily improve
the analytic results that can be obtained with a mixed order
one. However, first order formulations may be preferred in numerical
computations and in the analysis of the motion of isolated bodies
---notice, however, the local existence result for isolated bodies
consisting of dust (charged and uncharged) given in \cite{ChoFri06}
which is based on a mixed order formulation. In any case, to the best
of the author's knowledge, there is no frame formulation of the
Einstein field equation with is used in current numerical simulations
of the Einstein field equations. From our particular point of view,
the value of  first order formulations
of the Einstein-Euler system like the one being used here lies in
their amenability to a detailed analysis of
their linear and non-linear stability ---see
\cite{Reu99,AlhMenVal10}. Alternative discussions of the problem of
the well-posedness of the evolution equations of the
Einstein-Maxwell-Euler system and of magnetohydrodynamics can be found
in, for example, \cite{Put91,Fri74,Ren11,Put98,Zen03}, and
\cite{ChoYor02}.

\medskip
There are several formulations of the evolution equations of the
Einstein-Maxwell-Euler system geared towards its use in numerical
computations. Without the aim of being exhaustive, we notice in
particular \cite{BauSha03,ShiSek05,EtiLiuSha10} ---a good point of
entry to the extensive literature in this respect is the review
\cite{Fon03}. A peculiarity of the systems given in
\cite{BauSha03,ShiSek05,EtiLiuSha10} is their \emph{explicit use of
the conditions of ideal magnetohydrodynamics} to construct their
system. Hence, they may require some modifications if one is to use
them to model more general types of electromagnetic fields. The
question of the hyperbolicity of these systems is by no means direct
and has to be analysed with certain amount of care. In particular the
reference \cite{BauSha03} makes use of an ADM formulation for the
\emph{Einstein part} of the system which is known to have problems
concerning hyperbolicity ---see e.g. \cite{Fri96}.  This problem has
been addressed in \cite{ShiSek05,EtiLiuSha10} by using a so-called
BSSN formulation for the geometric part ---see e.g. \cite{Alc08} for
further details. The BSSN equations are first order in time and second
order in the spatial coordinates and are known to satisfy, in the
vacuum case, a certain notion of hyperbolicity (\emph{strong
hyperbolicity}) under certain additional conditions ---see
e.g. \cite{GunGar06}. Local existence and uniqueness of solutions to
strong hyperbolic systems follows under some further assumptions
---see e.g. \cite{Ren08}. However, because of the mixed order nature of the
system, it is not directly obvious that this hyperbolicity property is
preserved if one extends the vacuum Einstein system to the
Einstein-Maxwell-Euler one despite the \emph{lower order coupling} of
gravity to the charged fluid. To the best of our knowledge this issue
has not been addressed in the literature.  A discussion on the caveats of this type
of reasoning can be found, for example, in Section 8.5 of \cite{Ren08}.

\medskip
It should be pointed out that the evolution equations discussed in the
present article are only suited to the discussion of unbounded
self-gravitating charged fluids. In order to accommodate an initial
boundary value problem, further structure needs to be verified in the
equations. For example, it is not known whether our evolution
equations are compatible with the class of  maximally
dissipative boundary conditions used in, say \cite{FriNag99}, to show the
well-posedness of the initial boundary problem for the vacuum Einstein
field equations. In any case, it is interesting to point out that the analysis of
\cite{FriNag99} makes use of a frame formalism similar to the one used
in the present article.

\medskip
Symmetric hyperbolicity, despite being a
  technical mathematical notion has a deep physical
  significance. Moreover, the question of the well-posedness of the evolution equations of the
Einstein-Maxwell-Euler system is a problem that touches
upon many aspects of current theoretical and numerical analysis of
physical phenomena. Particular examples of these systems are given by
by accretion disks around compact objects. Due to the complexity of
the underlying equations, the analysis of the stability and the shape
of disks and flows is often addressed by numerical methods.  Further
applications of charged perfect fluids in General Relativity can be
found in, for example, \cite{Fon03,Zen03,BarMaaTsa07,Tsa05}. Numerical
simulations of magnetohydrodynamical systems with astrophysical
applications in view have been a topic of interest in recent years
---see for example
\cite{PalGarLehLie10,PalLehYos10,PalAndLehLieNei10,MoePalRezLehYosPol10,GiaRez07}.

\medskip
The present article is structured as follows: The tetrad formalism
used in this article is briefly reviewed in
Sec. \ref{Sec:CHANGENOTATION}. In
Sec. \ref{Sec:thebasiceq} we write and discuss the relativistic
equations describing a charged perfect fluid.  General remarks
concerning the reduction procedure to obtain suitable
evolution equations are given in Sec. \ref{Sec:redu}; the
resulting evolution  equations are discussed in
Sec. \ref{Sec:EvWeu} and the subsequent sections. The auxiliary field
describing the covariant derivative of the Faraday tensor is presented
in Sec. \ref{Sec:psi}. A summary of the evolution equations is
given in Sec. \ref{Section:Summary}. The case of an infinitely
conductive fluid (MHD) is briefly explored in
Sec. \ref{Subsection:MHD}. Some concluding remarks are given in
Sec. \ref{Sec:Colcusione}. Finally, in Appendix \ref{E:Aa}  we briefly
discuss certain issues concerning  homentropic flows and their
relation to the equation of state of the perfect fluid.

\section{Tetrad formalism}\label{Sec:CHANGENOTATION}
As mentioned in the introduction, our discussion of the
Einstein-Maxwell-Euler system will be based on a frame
formalism. The purpose of the present section is, mainly, to fix the
conventions to be used in the rest of the article.

\medskip
Let $(\mathcal{M},g_{\alpha\beta})$ denote a spacetime. Our discussion
will consider the two possible conventions for the signature of the
metric tensor
$g_{\alpha\beta}$. We introduce \emph{frame fields}
$\{e_{a}{}\}_{a=0,...,3}$ on the spacetime $\mathcal{M}$ satisfying
$g_{ab}\equiv g(e_{a}, e_{b})=\eta_{ab}=\epsilon \
\rm{diag}(1,-1,-1,-1)$  and $\epsilon=\pm 1$ ---see
\cite{MisThoWhe73,EllEls98}. We denote by $\left(\omega^{a}{}\right)$ the
corresponding \emph{dual basis} (cobasis)\footnote{Latin letters $a$ are the tetrad indices  while Greek letters $\alpha$ denotes
the tensorial character of each tensor (spacetime indices). Latin
letters $(i,j,k,q,p)$,  run in $\{1,2,3\}$,  denoting quantities in
the tetrad space.}.  The frame fields $e_a$ and the cobasis $\omega^a$
are expressed in terms of
a local coordinate basis as
\be\label{E:tetr1}
e_{a}=e_{a}{}^{\alpha}\partial_\alpha, \qquad \omega^a = \omega^a{}_\alpha \mbox{d}x^\alpha.
\ee
Thus,
\be\label{E:leggiadro1}
\omega^{a}{}_{\alpha}e_{b}{}^{\alpha}=
\delta^{a}{}_{b}, \qquad
\omega^{a}{}_{\alpha}e_{a}{}^{\beta}=\delta_{\alpha}{}^\beta.
\ee
The metric tensor and its inverse can be written as can be written in
terms of $\omega^a$ and $e_a$ as
\be\label{E:applica100}
g_{\alpha\beta}=\eta_{ab}\omega^{a}{}_\alpha\omega^{b}{}_\beta,\qquad
g^{\alpha \beta}=\eta^{ab}e_{a}{}^\alpha e_{b}{}^\beta.
\ee
The \emph{coefficients of the commutator of the elements of the
tetrad}, $D^{\phantom\ \phantom\ c}_{ ab}$, are defined by
\be\label{E:h2o1}
\left[e_{a},e_{b}\right]\equiv
D^{\phantom\ \phantom\ c}_{
ab}e_{c}=\left(e_{a}{}^{\alpha}\partial_{\alpha}
e_{b}{}^{\beta}-
e_{b}{}^{\alpha}\partial_{\alpha}e_{a}{}^{\beta}\right)\partial_{\beta}.
\ee
The \emph{connection components} (Ricci rotation coefficients),
$\g{a}_{b\ c}$,  of  the tetrad $ e_{a}$ are defined by the relations
\be\label{E:trconn}
\nabla_{a}e_{b}=\g{c}_{a\ b}e_c,
\qquad\nabla_{a}\omega^{b}=-\g{b}_{a\ c}\omega^c.
\ee
In particular, since
$e_{a}\left(\eta_{bc}\right)=0$, one has the symmetry
$\Gamma_{a(bc)}=0$. The \emph{Riemann tensor}, $R^{a}_{\phantom\
{bcd}}$, satisfying the \emph{Ricci identity}
\be\label{E:elastico}
R^{a}_{\phantom\
{bcd}}\omega_{a}=\nabla_{c}
\nabla_{{d}}\omega_{a}-\nabla_{{d}}\nabla_{c}\omega_{a}
-\nabla_{\left[c,{d}\right]} \omega_{b},
\ee
is given in terms of the
connection coefficients $\g{a}_{b\ c}$ by
\be\label{E:D41}
R^{a}_{\phantom  \
{{bcd}}}=e_c\left(\g{a}_{
{{d\ b}}}\right)-e_d\left(\g{a}_{
{{c\ b}}}\right)-\g{a}_{e\ b}(\g{e}_{c\ d}-\g{e}_{d\ c})+\g{a}_{c\ e}\g{e}_{d\ b}-\g{a}_{d\ e}\g{e}_{c\ b}.
\ee
In the sequel we will make use of the following decomposition of the  Riemann tensor
\be\label{E:decom}
R_{abcd}=C_{abcd}+ \left\{g_{{a[c}}S_{{d]b}}-g_{{b[c}}S_{{d]a}}\right\},
\ee
in terms of the \emph{Schouten tensor} $S_{ab}$
\be\label{E:S}
S_{ab}\equiv R_{ab}-\frac{1}{6}R g_{ab},
\ee
and the \emph{Weyl tensor}
\be
C_{abcd}\equiv
R_{abcd}+(g_{{a[c}}R_{{d]b}}-g_{{b[c}}R_{{d]a}})+\frac{1}{3}\ R\
g_{{a[c}}g_{{d]b}}.
\ee
In the previous expressions $R_{ab}\equiv R^{c}_{\phantom\ {acb}}$
denotes the \emph{Ricci tensor} of $g_{\alpha\beta}$, and  $R\equiv
g^{ab}R_{ab}$ its \emph{Ricci scalar}. The components of the curvature tensor satisfy the \emph{Bianchi identity}
\be\label{E:CH3COOH}
R^{a}_{\phantom\ b[{cd;e}]}=0.
\ee
Now, from the contracted Bianchi identities  for  the \emph{Einstein
  tensor}, $G_{ab}\equiv R_{ab}-\frac{1}{2}g_{ab} R$, satisfying
\begin{equation}\label{E:eqcr}
\nabla^{a}G_{ab}=0,
\end{equation}
we infer that
\be\label{E:hand}
F_{{abc}}\equiv  \nabla_{{d}}F^{{d}}_{\phantom\ {abc}}=0,
\ee
where $F_{abcd}$ denotes the so-called \emph{Friedrich tensor}
\be\label{E:F1}
F_{abcd}\equiv C_{abcd}-g_{{a[c}}S_{{d]b}}.
\ee
Taking the Hodge dual of Eq.\il(\ref{E:hand}) with respect to the
index pair $({c,d})$, we obtain an equation similar to (\ref{E:hand})
for the tensor $\tilde{F}_{abcd}$ defined by
\be\label{E:F1STAR}
\tilde{F}_{abcd}\equiv C^*_{abcd}+\frac{1}{2}S_{{pb}}\epsilon^{{p}}_{\phantom\ {acd}},
\ee
where $\epsilon_{abcd}$ denotes the components of the
\emph{completely antisymmetric Levi Civita tensor} with respect to the frame
$e_a$ and $C^*_{abcd}\equiv\frac{1}{2}C_{{abef}}\epsilon^{{ef}}_{\phantom\ \phantom\ cd}$.

\section{Ideal magnetohydrodynamics (MHD) in curved
  spacetime}\label{Sec:thebasiceq}
In this section we introduce the basic equations describing a
relativistic charged
perfect fluid coupled gravity ---the Einstein-Maxwell-Euler
system.  The notation and
conventions used are based on those of \cite{Fri98c} and
\cite{BarMaaTsa07,Tsa05}.

\medskip
We start by considering the following \emph{Einstein equations} \be
G_{a b}=\kappa T_{ab}, \ee with an \emph{energy-momentum tensor},
$T_{ab}$, which can be split as $T_{ab}\equiv T^{\rm{f}}_{a b}+
T^{\rm{em}}_{a b}$ where
\be
\label{E:Tm} T^{\rm{f}}_{a b}=(\rho +p) U_{a}
U_{b}-\epsilon\ p g_{a b}
\ee
is the energy-momentum tensor for an
\emph{ideal fluid}, while $\rho$ and $p$ are, respectively, the \emph{total
energy density} and \emph{pressure} as measured by an observer moving
with the fluid.  The time-like vector field (\emph{flow vector}) $U^a$
denotes the normalized future directed 4-velocity of the fluid. It
satisfies \be\label{E:Cos} U^a U_a=\epsilon.  \ee Associated to $U^a$
we introduce the \emph{projection tensor}
\be\label{E:h}
h_{ab}\equiv g_{ab}-\epsilon U_a U_b,
\ee
projecting onto the three dimensional subspace orthogonal to
$U^a$. Indices are raised and lowered with $g_{ab}$. Thus, one has
that $h^a{}_{b}=\delta^a{}_{b}-\epsilon U^a U_b$,
$h^b{}_ah^a{}_c=h^b{}_c$, and $h^a{}_bU_a=0$.

\subsection{The electromagnetic energy-momentum tensor}
The tensor $T^{\rm{em}}_{a b}$ denotes the energy momentum tensor of
an electromagnetic field: \be\label{E:ff} T^{\rm{em}}_{a
b}=-\epsilon\left (F_{a c}F^{\phantom\ c}_{b}-\frac{1}{4} F_{c d} F^{c
d} g_{ab}\right), \ee where $F_{ab}$ is the \emph{electromagnetic
field (Faraday) tensor}. The latter can be split in its \emph{electric
part}, $E_{a}\equiv F_{ab}U^{b}$, and its \emph{magnetic part},
$B^{a}\equiv\frac{1}{2}\epsilon^{abcd}U_{b}F_{cd}$, with respect to
the flow. Indeed, one has that \be F_{ab}=\epsilon( 2 E_{[a}U_{b]}
-\epsilon_{abcd}B^{c}U^{d}).  \ee Alternatively, the one can write
\be\label{E:disco} F_{ab}=\epsilon\left(E_{ab}-{}^*B_{ab}\right), \ee
where \be\label{E:dan} E_{ab}\equiv2E_{[a}U_{b]},
\quad{}^*E_{ab}\equiv\epsilon_{abcd}E^c U^d,\quad
B_{ab}\equiv2B_{[a}U_{b]}, \quad{}^*B_{ab}\equiv\epsilon_{abcd}B^c
U^d.  \ee One can readily verify the properties

\be B_{ab}U^b=\epsilon B_a,\quad E_{ab}U^b=\epsilon E_a,\quad
{}^*B_{ab}U^b={}^*E_{ab}U^b=0, \ee where, as before, ${}^*$ denotes
the Hodge dual operator. In particular, one has
that,${}^{**}B_{ab}=-B_{ab},$ so that
${}^*F_{ab}=\epsilon\left({}^*E_{ab}-{}^{**}B_{ab}\right)=
\epsilon\left({}^{*}E_{ab}+B_{ab}\right),$ with \be
B_a={}^*F_{ab}U^b,\quad{}^*F_{ab}=\frac{1}{2}\epsilon_{abcd}F^{cd},\quad{}^{**}F_{ab}=-F_{ab}.
\ee From the decomposition into electric and magnetic parts, one can
readily verify that the Faraday tensor has six independent components
as these vectors are both spatial ---that is,
$E_{a}U^a=B_aU^a=0$. Using the decomposition into electric and
magnetic parts, the electromagnetic energy-momentum tensor,
Eq.\il(\ref{E:ff}), can be written as
\be\label{E:ff2}
T^{\rm{em}}_{a b}\equiv-\frac{\epsilon}{2} U_aU_b (E^2+B^2)+\frac{h_{ab}}{6}(E^2+B^2)+P_{ab}-2\epsilon \mathcal{G}_{(a}U_{b)},
\ee
where we have written $E^2\equiv E_aE^a$ and $B^2 \equiv B^aB_a$, and $P_{ab}$ denotes the symmetric, trace-free tensor given by
\be
P_{ab}=P_{(ab)}\equiv\frac{h_{ab}}{3}(E^2+B^2)-(E_aE_b+B_aB_b).
\ee
Furthermore,
\be
\mathcal{G}_{a}\equiv\epsilon_{auvd}E^uB^vU^d,
\ee
denotes the \emph{Poynting vector}. Alternatively, one can
rewrite Eq.\il(\ref{E:ff2}) in the form
\be\label{E:fying}
T^{\rm{em}}_{ab}=\frac{g_{ab}}{2}(E^2+B^2)-(E_aE_b+B_aB_b)-2\epsilon \mathcal{G}_{(a}U_{b)}-\epsilon U_aU_b(E^2+B^2).
\ee

\subsection{The Maxwell equations}
The Maxwell equations are given by
\be\label{E:MW}
\nabla_{[a}F_{b c]}=0,\quad\nabla^a F_{ab}=\epsilon J_b.
\ee
Using Eq.\il(\ref{E:disco}) one can rewrite them as
\be\label{E:dic}
\nabla_{[a}\left(E_{bc]}-{}^*B_{bc]}\right)=0,
\quad \nabla^{a}\left(E_{ab}-{}^*B_{ab}\right)=J_b.
\ee
The homogeneous Maxwell equation can be rewritten in
terms of the Hodge dual of $F_{ab}$ as
\be\label{E:tric}
\nabla^a{}^*F_{ab}=\epsilon\nabla^a\left({}^*E_{ab}+B_{ab}\right)=0.
\ee
Eq.\il(\ref{E:dic}) can be further expanded to yield
\be\label{E:treert}
\left(\nabla^bE_b\right)U_a-\dot{E}_a+\nabla^b\epsilon_{abcd} B^c U^d+2E_{[b}\nabla^bU_{a]}=J_a,
\ee
where $\dot{E}\equiv U^a\nabla_a E_b$ stands for the covariant
derivative of $E_a$ along the flow. For Eq.\il(\ref{E:tric}) a
similar computation gives
\be\label{E:ancor}
\nabla^a\left(\epsilon_{abcd}E^c U^d+2 B_{[a}U_{b]}\right)=0.
\ee
Projecting Eq.\il(\ref{E:treert}) along the directions parallel and
orthogonal to the flow vector $U^b$ one obtains the  two equations
\begin{eqnarray}
&& \epsilon \nabla^aE_a+E_b\dot{U}^b-\epsilon_{abcd}U^a B^b\nabla^c U^d=U^bJ_b,\label{E:UE1} \\\nonumber
&& \epsilon U_e\epsilon_{abcd}U^b B^c \nabla^a U^d-\left(\dot{E}_e+\epsilon U_e E^b \dot{U}_b\right)-\nabla^a(\epsilon_{aecd}B^c U^d)\\
&&+h^b{}_e E_a \nabla^aU_b-E_e \nabla^aU_a=h^b{}_e J_b,\label{E:UE2}
\end{eqnarray}
Projecting Eq.\il(\ref{E:ancor}) along the directions parallel and
orthogonal to $U^a$ one obtains
\begin{eqnarray}
&&\epsilon_{abcd}U^bE^c\nabla^a U^d+\epsilon \nabla^aB_a-U^b \dot{B}_b=0,\label{E:Ub1} \\
&& \epsilon_{abcd}h^b{}_f\nabla^a(E^cU^d)+B_a h^b{}_f\nabla^a U_b-h^b{}_f \dot{B}_b-B_f \nabla^a U_a=0,\label{E:Ub2}
\end{eqnarray}
where it has been used that
$\nabla_{a}(E^bU_b)=E_b\nabla_aU^b+U_b\nabla_aE^b=0$.

\medskip
The electromagnetic current vector $J^{a}$ will be split with respect
to the flow vector as
\be\label{E:J}
J^a=\rc U^a +j^a,
\ee
where $\rho_{\ti{C}}$ denotes the \emph{charge density} and $j^a$ is
the \emph{orthogonally projected conduction current}. Using the
decomposition Eq.\il(\ref{E:J}) in Eqs.(\ref{E:UE1}) and
(\ref{E:UE2}), one obtains the following propagation equations for the
electric and magnetic parts of the Faraday tensor:
\bea\label{E:UE2a}
\dot{E}_{\langle f \rangle}&=&-2E^a h_{f[a}\nabla_{b]} U^b-\epsilon_{abcd}h^b{}_f\nabla^a(B^c U^d)-j_f,
\\
\label{E:Ub2a}
\dot{B}_{\langle f \rangle}&=&-2B^a h_{f[a}\nabla_{b]} U^b+\epsilon_{abcd}h^b{}_f\nabla^a(E^cU^d),
\eea
and from Eqs.(\ref{E:Ub1}) and (\ref{E:Ub2}) the constraint equations
\bea\label{E:Ub1a}
\epsilon D^aB_a&=&-\epsilon_{abcd}U^bE^c\nabla^a U^d,
\\
\label{E:UE1a}
\epsilon D^aE_a&=&\epsilon_{abcd}U^a B^b\nabla^c U^d+\epsilon \rho_{\ti{C}},
\eea
where $\dot{w}_{\langle a \rangle}\equiv h_a{}^b \dot{w}_b$ denotes
the directional covariant derivative along the flow vector
(\emph{Fermi derivative}) and $D_aw_b\equiv h_a{}^u
h_b{}^v\nabla_uw_v$ is the orthogonally projected covariant derivative
of a vector. We note that
$\epsilon_{abcd}h_f{}^b\nabla^a(X^cU^d)=-\mbox{curl} X_f+\epsilon
\epsilon_{afcd}U^a X^c \dot{U}^d$, where $\mbox{curl}
X_f\equiv\epsilon_{facd}U^d \nabla^a B^c$.

\medskip
In the sequel, it will be necessary to specify the form of the
conduction current, $j^a$. Consistent with \emph{Ohm's law}, we assume
a linear relation between the conduction current $j^a$ and
the electric field. More precisely, we set
\be
j^a=\sigma^{ab}E_b,
\ee
where $\sigma^{ab}$ denotes the \emph{conductivity} of the fluid
(plasma). We will restrict our attention to isotropic fluids for which
$\sigma^{ab}=\sigma g^{ab}$, so that Eq.\il(\ref{E:J}) becomes
\be\label{E:lg}
J^a=\rc U^a +\sigma E^a,
\ee
with $\sigma$ the \emph{electrical conductivity coefficient}. Ideal
MHD is characterised by the condition, $\sigma\rightarrow\infty$
(ideal conductive plasma). It follows from Eq.\il(\ref{E:lg}) that for
this to be the case one requires the constraint $E_a=0$.

\subsection{The fluid equations}
Form  the conservation of the energy-momentum tensor
\be\label{:B}
\nabla^a T_{ab}=0,
\ee
and using Eqs.\il(\ref{E:Tm}) and \il(\ref{E:ff}) and the Maxwell
equation Eq.\il(\ref{E:MW}) one obtains
\bea
&& \nabla^{a}T^{\rm{em}}_{ab}=-\epsilon\left(\nabla^a F_{ac}\right)F_{b}^{\phantom\ c}, \\
&& \nabla^{a}T^{\rm{f}}_{ab}=U_bU_a\nabla^a (p+\rho)+(p+\rho)\left[U_b(\nabla^aU_a)+U_a \nabla^a U_b\right]-\epsilon\nabla_b p.
\eea
Therefore, Eq.\il(\ref{:B}) reads:
\be\label{E:Tor}
U_bU_a\nabla^a (p+\rho)+(p+\rho)\left[U_b(\nabla^aU_a)+U_a \nabla^a U_b\right]-\epsilon\nabla_b p-\epsilon\left(\nabla^a F_{ac}\right)F_{b}^{\phantom\ c}=0.
\ee
In what follows, we will consider the projections of (\ref{E:Tor})
along the directions parallel and orthogonal to the flow lines of the
fluid. Contracting Eq. (\ref{E:Tor}) with $\epsilon U^b$ we obtain the
conservation equation
\be\label{Eq:conservazione}
U_a\nabla^a\rho+(p+\rho)\nabla^aU_a- U^bF_{b}^{\phantom\ c}(\nabla^aF_{ac})=0,
\ee
where the constraint (\ref{E:Cos}) has been used. For an ideally
conducting fluid, where $E_a=F_{ab}U^{b}=0$, the last term of
Eq.\il(\ref{Eq:conservazione}) is identically zero and the
electromagnetic field does not have a direct effect on the
conservation equation along the flow lines. Contracting (\ref{E:Tor}) with the projector $h^{bc}$ one obtains the \emph{Euler equation}
\be\label{E:qwety}
(p+\rho)U^a\nabla_aU^c-\epsilon h^{bc}\nabla_b p-\epsilon(\nabla^aF_{ad})F^{\phantom\ d}_b h^{bc}=0.
\ee
The later equation can also be written as
\be\label{E:qwety2}
(p+\rho)U_a\nabla^aU_b+(U_bU^d\nabla_d-\epsilon \nabla_b) p-\epsilon(\nabla^aF_{ad})\left(F^{\phantom\ d}_b-\epsilon F^{ed}U_e U_b\right)=0.
\ee
In the ideal MHD case the last term of Eq.\il(\ref{E:qwety2}) is identically zero so that Eq.\il(\ref{E:qwety}) reduces to
\be
(p+\rho)U^a\nabla_aU^c-\epsilon h^{bc}\nabla_b p-\epsilon(\nabla^aF_{ad})F^{cd}=0.
\ee

\subsection{Thermodynamic considerations}

Here we consider a \emph{one species particle fluid} (simple
fluid). We denote by $n,\,s,\, T$ the \emph{particle number density},
the \emph{entropy per particle} and the \emph{absolute temperature},
as measured by comoving observers. We also introduce the \emph{volume
per particle}, $v$, and the \emph{energy per particle}, $e$, via
\be
v\equiv\frac{1}{n}, \quad e \equiv \frac{\rho}{n}.
\ee
In terms of these variables the first law of Thermodynamics,
$\mbox{d}e=-p\mbox{d}v+T\mbox{d}s$, takes the form
 \be\label{E:1law}
\mbox{d}\rho=\frac{p+\rho}{n} \mbox{d}n +n T \mbox{d}s.
\ee
Assuming an equation of state of the form $\rho=f(n,s)\geq0$, one obtains from
 Eq.\il(\ref{E:1law}) that
\be\label{E:the}
p(n,s)=n\left(\frac{\partial \rho}{\partial n}\right)_{s}-\rho(n,s),
\quad T(n,s)=\frac{1}{\rho}\left(\frac{\partial \rho}{\partial s}\right)_n.
\ee
Assuming that $\partial p/\partial \rho>0$ we define the \emph{speed
of sound}, $\nu_s=\nu_s(n,s)$, by
\be\label{E:vs}
\nu_s^2\equiv\left(\frac{\partial p}{\partial \rho}\right)_s=\frac{n}{\rho+p}\frac{\partial p}{\partial n}>0.
\ee

\medskip
Since we are not considering particle annihilation or creation we
consider the equation of conservation of particle number:
\be\label{E:then}
U^a\nabla_a n+n \nabla_a U^a=0.
\ee
Combining this equation with Eqs.\il(\ref{Eq:conservazione}) and \il(\ref{E:1law}) we obtain
\be\label{E:fors}
U^a\nabla_a s=\frac{1}{nT}U^b F_{b}^{\phantom\ c}\nabla^a F_{ac}.
\ee
In the case of an infinitely conducting plasma, where the last term of
Eq. (\ref{E:1law}) vanishes, Eq.\il(\ref{E:fors}) describes an
adiabatic flow ---that is, $U^a\nabla_as=0$, so that the entropy per
particle is conserved along the flow lines. A particular case of
interest is when $s$ is a constant of both space and time.  In this
case the equation of state can be given in the form $p=p(\rho)$. A
further discussion on  homentropic flows and barotropic equations of
state is provided in Appendix\il\ref{E:Aa}.

\section{The hyperbolic reduction procedure}\label{Sec:redu}

\subsection{General considerations}
Following  \cite{Fri98c},  we introduce the notation
\be
N^{a}\equiv\delta_0{}^a,\quad
N\equiv N^{a}e_a=e_0.
\ee
 For a given tensor any contraction with $N$ will be denoted by
replacing the corresponding index by $N$, and the projection with
respect to $h_a{}^b$ will be indicated by a prime. Accordingly, for a tensor
$T_{abc}$ one writes
\be
T'_{aNb}\equiv T_{mpq}h_{a}{}^{m}N^p h_{b}{}^q.
\ee
Introducing the notation $\epsilon_{abc}=\epsilon'_{Nabc}$, where $\epsilon_{0123}=1$, we obtain the decomposition
\be
\epsilon_{abcd}=2 \epsilon\left(N_{[a}\epsilon_{b]cd}-\epsilon_{ab[c}N_{d]}\right).
\ee
Given a spatial tensor $T_{a_1\cdots a_p}=T'_{a_1 \cdots a_p}$ we define the \emph{spatial covariant derivative}
\be
\mathcal{D}_aT_{a_1\cdots a_p}=h_{a}{}^b h_{a_1}{}^{b_1}\cdots
h_{a_p}{}^{b_p} \nabla_b T_{b_1 \cdots b_p}.
\ee
One has, in particular, that $\mathcal{D}_ah_{bc}=\dd_a\epsilon_{bcd}=0$.

\medskip
For convenience, we introduce the tensors
\be\label{E:dozero}
a^a \equiv N^{b}\nabla_{b}N^a, \quad \chi_{ab} \equiv h_{a}^c\nabla_c N_b, \quad  \chi\equiv h^{ab}\chi_{ab}.
\ee
In terms of these we can write
\be\label{E:rezero}
\nabla_aN^b=\epsilon N_aa^b +\chi_a{}^b,\quad a^a=h_b{}^a\Gamma^b_{00},\quad\chi_{ab}=-\epsilon h_a{}^c h_{b}{}^d\Gamma^0{}_{cd}.
\ee

\medskip
Key for our subsequent discussion will be the decomposition of the
Weyl tensor on terms of its \emph{electric}, $\hat{E}_{ab}\equiv
C'_{NaNb}$, and \emph{magnetic} parts, $\hat{B}_{ab}={C^{*}}'_{NaNb}$,
with respect to $N_a$  given by
\be\label{E:Cabcd}
C_{abcd}=2 \epsilon(l_{b[c}\hat{E}_{d]a}-l_{a[c}\hat{E}_{d]b})-2(N_{[c}\hat{B}_{d]p}\epsilon^p_{\phantom\ ab}+N_{[a}\hat{B}_{b]p}\epsilon^p_{\phantom\ cd})
\ee
where \(l_{ab}\equiv h_{ab}-\epsilon N_a N_b\). From the  Bianchi identity Eq.\il(\ref{E:hand}) for the tensor $F_{abcd}$ we obtain the  decomposition
\be
F_{abc}=N_a\left(F'_{NbN}N_c-F'_{NcN}N_b\right)-2\epsilon F'_{aN[b}N_{c]}+\epsilon N_{a}F'_{N bc}+F'_{abc},
\ee
from where it follows that
\bea
\nonumber
&&\epsilon F'_{aNb}=\mathcal{L}_{N}F'_{NaNb}+\epsilon \dd^cF'_{caNb}-a^c\left(F'_{Nacb}+F'_{caNb}\right)+\epsilon a_a F'_{NNNb}
\\
\label{E:decores}
&& \hspace{2cm} -\epsilon \chi^{cd}F'_{cadb}-\chi_a^{\phantom\ c}F'_{NcNb}-\chi_b^{\phantom\ c}F'_{NaNc}+\chi^c_{\phantom\ a} F'_{cNNb}+\chi F'_{NaNb}.
\eea

\subsection{Detailed analysis of the evolution equations}\label{Sec:EvWeu}
In the sequel we make the following specific choice of timelike frame vector:
\[
N=e_0=U,\quad N_a=U_a.
\]

Using the decomposition given by Eq.\il(\ref{E:F1}) we compute the
following components of the tensors $F_{abcd}$ and
$\tilde{F}_{abcd}$:
\bea
\tilde{F}'_{NNNa}&=&0, \quad F'_{NNNa}=-\frac{\epsilon \kappa}{2} T^{\rm{em}}_{ec}U^c h^e{}_a,
\\
\tilde{F}'_{NaNb}&=&\hat{B}_{ab}, \qquad F'_{NaNb}=\hat{E}_{ab}+\frac{\kappa\rho}{6} h_{ab}-\frac{\kappa\epsilon}{2} T^{\rm{em}}_{dc}h^d{}_b h^c{}_a,
\\
\tilde{F}'_{NNab}&=&\frac{\kappa}{2}\left(T^{\rm{em}}_{pv}U^v\right)\epsilon{}^p_{uab}U^u,
\quad F'_{NNab}=0,
\\
\tilde{F}'_{aNNb}&=&-\hat{B}_{ab}+\frac{\kappa}{2}\left(T^{\rm{em}}_{pv}U^v\right)\epsilon{}^p_{aub}U^u,
\\
F'_{aNNb}&=&-\hat{E}_{ab}+\frac{\kappa}{2}h_{ab}\left[\left(\frac{2}{3}\rho+p\right)+T^{\rm{em}}_{uv}U^vU^u\right],
\\
\tilde{F}'_{Nabc}&=&\epsilon \hat{E}_{ap}\epsilon{}^p_{bc}-\frac{\kappa \rho \epsilon}{6}\epsilon_{aubc}U^u+
\frac{\kappa}{2}\left(T^{\rm{em}}_{pv}h^v{}_a\right)\epsilon^p{}_{ubc}U^u, \quad F'_{Nabc}=-\epsilon \hat{B}_{ap}\epsilon^p{}_{bc},
\\
\tilde{F}'_{aNbc}&=&-\epsilon \hat{E}_{ap}\epsilon^p{}_{bc}+\frac{\epsilon \kappa}{2}\left(\frac{2}{3}\rho+p\right)U_p\epsilon^p{}_{abc}+\frac{\kappa}{2}\left(T^{\rm{em}}_{up}U^u\right)
\epsilon^p{}_{dvf}h^d{}_ah^v{}_bh^f{}_c,
\\
F'_{aNbc}&=&\hat{B}_{ap}\epsilon^p{}_{bc}-\frac{\kappa}{2}\left[h_{ab}\left(T^{\rm{em}}_{fu}U^uh^f{}_c\right)-
h_{ac}\left(T^{\rm{em}}_{fu}U^uh^f{}_b\right)\right],
\\
\tilde{F}'_{abNc}&=&-2\epsilon \hat{E}_{p[b}\epsilon^{\phantom\ p}_{a]\phantom\ c}-\frac{\kappa \rho \epsilon }{6} \epsilon_{bavc}U^v+\frac{\kappa}{2}T^{\rm{em}}_{pu}h^u{}_b\epsilon^p{}_{avc}U^v,
\\
F'_{abNc}&=&-\epsilon \hat{B}_{cp}\epsilon^p{}_{ab}+\frac{\kappa}{2} h_{ac}T^{\rm{em}}_{uv}U^uh^v{}_b,
\\
\tilde{F}'_{abcd}&=&-\hat{B}_{pq}\epsilon^p{}_{ab}\epsilon^q{}_{cd}-\frac{\epsilon\rho \kappa}{6} \epsilon_{pu ef}h^p{}_bh^u{}_a h^e{}_c h^f{}_d+\frac{\kappa}{2}\left(T^{\rm{em}}_{pv}h^v{}_b\right)\epsilon^p{}_{uef}h^u{}_ah^e{}_c h^f{}_d,
\\
F'_{abcd}&=&2 \epsilon\left[l_{b[c}\hat{E}_{d]a}-l_{a[c}\hat{E}_{d]b}\right]+\frac{\epsilon \kappa \rho}{3}h_{c[a}h_{b]d}-\kappa
T^{\rm{em}}_{uv}h^v{}_bh_{c[a}h_{d]}{}^u.
\eea
Moreover, Eq.\il(\ref{E:S}) becomes,
\be\label{E:sab}
S_{ab}=S^{\rm{f}}_{ab}+S^{\rm{em}}_{ab},
\ee
where
\be
S^{\rm{f}}_{ab}\equiv \kappa\left[T^{\rm{f}}_{ab}-\frac{1}{3}g_{ab}g^{cd} T^{\rm{f}}_{cd}\right]=\kappa\left[\left(\frac{2}{3}\rho +p\right)U_a U_b-\frac{\epsilon  \rho}{3} h_{ab}\right],
\ee
and
\be\label{E:congreg}
S^{\rm{em}}_{ab}\equiv \kappa T^{\rm{em}}_{ab}= -\epsilon \kappa\left (F_{a c}F^{d c}g_{d b}-\frac{1}{4} F_{c d} F^{c d} g_{ab}\right).
\ee
Using  Eq.\il(\ref{E:sab}) and Eq.\il(\ref{E:Cabcd}) in Eq.\il(\ref{E:decom}), the Riemann curvature tensor $R_{abcd}$, can be decomposed in the form
\bea\label{E:recie}
R_{abcd}&=&R_{abcd}^{\ti{W}}+R_{abcd}^{\ti{m}}+R_{abcd}^{\ti{em}},
\eea
where
\bea
\nonumber
R_{abcd}^{\ti{em}}\equiv && \kappa\bigg((E^2+B^2)\left[\frac{1}{2}\left(g_{a[c}g_{d]b}-g_{b[c}g_{d]a}\right)-2\epsilon U_{[b}g_{a][c}U_{d]}\right]-2E_{[b}g_{a][c}E_{d]}+2B_{[b}g_{a][c}B_{d]}\\
&&+2\epsilon \left[U_{[a}g_{b][c}\mathcal{G}_{d]}-\mathcal{G}_{[a}g_{b][c}U_{d]}\right]\bigg),
\eea
and
\bea
&& R_{abcd}^{\ti{W}}\equiv2 \epsilon(l_{b[c}\hat{E}_{d]a}-l_{a[c}\hat{E}_{d]b})-2(U_{[c}\hat{B}_{d]p}\epsilon^p_{\phantom\ ab}+U_{[a}\hat{B}_{b]p}\epsilon^p_{\phantom\ cd}), \\
&& R_{abcd}^{\ti{m}}=\frac{1}{3} \epsilon \kappa \rho \left(l_{b[c}h_{d]a}-l_{a[c}h_{d]b}\right)+\kappa p\left(h_{a[c}U_{d]}U_b-h_{b[c}U_{d]}U_a\right).
\eea
Finally, from the decomposition Eq.\il(\ref{E:decores}) we identify
the tensors
\[
\epsilon \tilde{F}'_{(a|N|b)}, \qquad \epsilon \left[F'_{(a|N|b)}-\frac{1}{2}h_{ab}h^{uv}F'_{uNv}\right], 
\]
As it will be shown in the next subsections, the evolution equations
for the electric and magnetic parts of the Weyl tensor will be
extracted from these tensors. The essential difficulty in our analysis
resides in a satisfactory treatment of these equations.

\subsubsection{Evolution equation for $\hat{B}_{ab}$}
From Eqs.\il(\ref{E:decores}) and \il(\ref{E:hand}) we obtain the
following evolution equation for the magnetic part of the Weyl tensor
\bea\nonumber
0=\epsilon \tilde{F}'_{(a|N|b)}&=&\mathcal{L}_U \hat{B}_{ab}-D_d \hat{E}_{c ( a }\epsilon_{b)}{}^{dc}+2\epsilon a_c \epsilon^{cd}{}_{( a}\hat{E}_{b)d}-\chi^c_{\phantom\ (a}\hat{B}_{b) c}
\\\label{E:sa}
&&\hspace{1cm} -2 \chi_{(a}{}^{c}\hat{B}_{b) c}+\chi \hat{B}_{ab}-\epsilon \chi_{cd}\hat{B}_{pq}\epsilon^{pc}{}_{(a}\epsilon^{dq}{}_{b)}+\epsilon \tilde{F}^{'\rm{em}}_{(a|N|b)},
\eea
where
\be\label{E:1utri} \epsilon \tilde{F}^{'\rm{em}}_{(a|N|b)}\equiv
\frac{\kappa\epsilon}{2}\mathcal{D}^c\left(\Tem_{up}
\epsilon^p{}_{cv(b}h^u_{a)}U^v\right)-\frac{\epsilon \kappa}{2}
\chi^{cd}\left(\Tem_{pv}h^v{}_{(a}h^f{}_{b)}\right)\epsilon^p{}_{uef}h^u{}_{c}h^e{}_{d}
+\frac{\kappa}{2}\epsilon^p{}_{cu(b}\chi^c_{\phantom\
a)}U^u\left(\Tem_{pv}U^v\right).  \ee In vacuum, Eq. (\ref{E:sa})
together with its electric counterpart provide hyperbolic evolution
equations for the electric and magnetic parts of the Weyl tensor
independently of the gauge choice. This is not automatically the case
in the presence of matter. Using Eq.\il(\ref{E:congreg}) in
Eq.\il(\ref{E:1utri}) we find that the electromagnetic contribution to
the evolution equation of $\hat{B}_{ab}$ is given by
\bea\nonumber \epsilon \tilde{F}^{'\rm{em}}_{(a|N|b)}&\equiv&
\frac{-\epsilon \kappa}{2}\left(\epsilon\mathcal{D}^c\left[U^v
\epsilon^p{}_{cv(b}\left(
h^u{}_{a)}F_{uq}F_{p}{}^{q}-\frac{1}{4}h_{a)p}F_{qs}F^{qs}\right)\right]\right.
\\ &&\left.-\epsilon\chi^{ue}
\epsilon^p{}_{uef}\left(h^f{}_{(b}h^v{}_{a)}F_{vc}F_p^{\phantom\
c}-\frac{1}{4}h^f{}_{(b}h_{a) p}F_{qs}F^{qs}\right)-U^u
\epsilon^p{}_{cu(b}\chi^c{}_{a)}E_dF_{p}{}^{d}\right).  \eea Notice
that this last expression contains derivatives of the Faraday tensor
which cannot be replaced by means of the Maxwell equations. These
derivatives enter into the principal part of the evolution equations
and destroy the hyperbolicity of the evolution equations for the
magnetic part of the Weyl tensor. In order to deal with this
difficulty and additional variable, representing the derivative of
the Faraday tensor needs to be introduced. This will be discussed in
subsection \ref{Sec:psi}.

\subsubsection{Evolution equation for $\hat{E}_{ab}$}
The evolution equation for the electric part of the Weyl tensor can be obtained from Eq.\il(\ref{E:decores}) as follows:
\bea\nonumber
0=\epsilon \left[F'_{(a|N|b)}-\frac{1}{2}h_{ab}h^{uv}F'_{uNv}\right]=\mathcal{L}_U \hat{E}_{ab}+D_c\hat{B}_{d(a}\epsilon_{b)}{}^{cd}-2 \epsilon a_c \epsilon^{cd}{}_{(a}\hat{B}_{b)  d}
-3 \chi_{(a}{}^{c}\hat{E}_{b) c}-2\chi^{c}{}_{(a}\hat{E}_{b) c}
\\
+h_{ab}\chi^{cd} \hat{E}_{cd}+2\chi \hat{E}_{ab}+\frac{\kappa}{2}(\rho+p)\left[\chi_{(ab)}-\frac{1}{3}\chi h_{ab}\right]+\epsilon \left[F_{(a|N|b)}^{'\rm{em}}-\frac{1}{2}h_{ab}h^{uv}F^{' \rm{em}}_{uNv}\right], \label{E:terete}
\eea
where
\bea
\nonumber
\epsilon \left[F^{'\rm{em}}_{(a|N|b)}-\frac{1}{2}h_{ab}h^{uv}F^{'\rm{em}}_{uNv}\right]
&=&\frac{1}{4}\epsilon \kappa \mathcal{L}_U \left(\Tem_{dc}h^d{}_{a}h^c{}_{b}\right)+\frac{\epsilon \kappa}{2}\left[h^{v}{}_{(a}\mathcal{D}_{b)}-\frac{1}{2}h_{ab}\mathcal{D}^v\right](\Tem_{uv}U^u)
\\
\nonumber
&&+\kappa \Tem_{uv}U^v\left[\frac{h_{ab}}{2}a^u-h^u{}_{(b}a_{a)}\right]
+\frac{\kappa}{2}\Tem_{uv}U^uU^v \left[\chi_{(ab)}-\frac{h_{ab}}{2}\chi\right]
\\
&&+\frac{\kappa\epsilon}{2}\left[2\chi_{(a}{}^{u}h^v{}_{b)}-\chi_{(a}{}^u h_{b)}{}^v-\frac{\chi^{uv}h_{ab}}{2}\right]\Tem_{uv}.
\eea
Finally, using Eq.\il(\ref{E:congreg}) we obtain the explicit expression
\bea \nonumber \epsilon
\left[F^{'\rm{em}}_{(a|N|b)}-\frac{h_{ab}h^{uv}}{2}F^{'\rm{em}}_{uNv}\right]
&=&-\frac{\kappa\mathcal{L}_U}{4}\left[h^q{}_{(a}h^c{}_{b)}F_{qf}F_c{}^{f}-
\frac{h_{ab}F_{qp}F^{qp}}{2}\right]
-\frac{\kappa}{2}\left[\frac{h_{ab}\mathcal{D}^v}{2}-h^v{}_{(a}\mathcal{D}_{b)}\right]
\left(E^c F_{vc}\right) \\ \nonumber &&-\kappa\epsilon E^c
F_{vc}\left(h^v{}_{(a}a_{b)}-\frac{h_{ab}}{2}a^v\right)-\frac{\kappa}{2}F_{ud}F_{v}{}^{d}\left[2\chi_{(a}{}^{v}h_{b)}{}^u-\chi^{v}{}_{(a}h_{b)}{}^u-\frac{1}{2}\chi^{uv}h_{ab}\right]
\\
&&+\frac{\kappa}{4}\chi_{(ab)}F_{qp}F^{qp}-\frac{\kappa\epsilon}{2}E^2
\chi_{(ab)}-\frac{1}{8}\chi h_{ab}F_{qp}F^{qp}+\frac{1}{4}\epsilon
\kappa \chi h_{ab}E^2.  \eea As in the previous subsection we observe
the presence of derivatives of the Faraday tensor which need to be
dealt with by the introduction of a new field if one is to preserve
the hyperbolicity of the equations.

\subsection{An auxiliary field}\label{Sec:psi}
As discussed previously, derivatives of the Faraday tensor appear in
the evolution equations implied by the Bianchi identity. This feature
of non-vacuum systems destroys, in principle, the hyperbolicity of the
evolution equations. In order to get around with this difficulty, it
is necessary to introduce the covariant derivative of the Faraday
tensor as a field variable. More precisely, let
\be\label{E:tono}
\psi_{abc} \equiv \nabla_a F_{bc}.
\ee
As a consequence of the definition and the Maxwell equations the
tensor $\psi_{abc}$ has the symmetries  $\psi_{abc}=\psi_{a[bc]}$, $\psi_{[abc]}=0$.
In order to construct a suitable equation for $\psi_{abc}$ we consider
the commutator of the covariant derivative $\nabla$ in the form
\be
\nabla_a \nabla_b F_{cd} - \nabla_b \nabla_a F_{cd} = -2 F_{e[c} R^e{}_{d]ab}.
\ee
Making use of the definition of $\psi_{abc}$ one obtains
\be\label{E:in}
\nabla_a \psi_{bcd} - \nabla_b \psi_{acd} = -2 F_{e[c} R^e{}_{d]ab},
\ee
where in the right hand side we use the expression for $R_{abcd}$
involving the its decomposition in terms of the Weyl tensor and the
matter fields. The tensorial expression (\ref{E:in}) provides the
required field equations.

\subsection{Propagation equations for the auxiliary field}

 In order to deduce from them suitable
evolution equations one proceeds by analogy to the case of the Faraday
tensor $F_{ab}$. Proceeding as in the case of Eqs. \il(\ref{E:MW})
and \il(\ref{E:tric}) we obtain the  following two equations for $\psi_{abc}$
\bea\label{E:star}
&& \nabla^b\psi_{adb}=2F^{e[b}R_{d]aeb}-\epsilon \nabla_a J_d, \\
&&\label{E:stars}
\nabla^a {}^*\psi_{cab}=\epsilon_b^{\phantom\ aud}F_{eu}R^e_{\phantom\ dac},
\eea
where  $R_{abcd}$ is given by Eq.\il(\ref{E:recie}). In the subsequent
discussion is convenient to introduce here the following notation
\be
\psi_{ayz}\equiv\Psi_{yz}= \Psi_{[yz]}.
\ee
That is,  the first index of $\psi_{abc}$ tensor is dropped. Because
of the $\psi_{abc}$ is antisymmetric with respect to the last pair of
indices, it is natural to introduce  its  \emph{electric} and \emph{magnetic} part respect to $U_a$ via
\be
\mathcal{E}_d\equiv \Psi_{dN}\equiv\psi_{adn}U^n, \qquad\mathcal{B}^u\equiv\frac{1}{2}\epsilon^{uvzt}U_v \Psi_{zt}, \label{PsiElectric-Magnetic}
\ee
respectively, so that are $\mathcal{E}_d$ and $\mathcal{B}_d$ are
spacelike vectors:
\be
\mathcal{E}_{a}U^a=\mathcal{B}_aU^a=0.
\ee
Accordingly, we write the tensor $\Psi$ as
\be\label{E:discobis}
\Psi_{ab}=\epsilon\left[\mathcal{E}_{ab}-{}^*\mathcal{B}_{ab}\right].
\ee
Furthermore, one has that
\be
^{*}\Psi_{ab}=
\epsilon\left[{}^{*}\mathcal{E}_{ab}+\mathcal{B}_{ab}\right],
\ee
where
\bea\label{E:danI}
\mathcal{E}_{ab}\equiv2\mathcal{E}_{[a}U_{b]},
\qquad{}^*\mathcal{E}_{ab}\equiv\epsilon_{abcd}\mathcal{E}^c U^d,
\\
\mathcal{B}_{ab}\equiv2\mathcal{B}_{[a}U_{b]}, \qquad{}^*\mathcal{B}_{ab}\equiv\epsilon_{abcd}\mathcal{B}^c U^d.
\eea

\medskip
Continuing the analogy with the Maxwell equations and the Faraday
tensor $F_{ab}$, it follows that Eqs.\il(\ref{E:star})  and Eq.\il(\ref{E:stars}) read
\bea
&& \label{E:treertb}
\dot{\mathcal{E}}_d-\left(\nabla^b\mathcal{E}_b\right)U_d-\nabla^b\epsilon_{dbcv}
\mathcal{B}^c U^v-2\mathcal{E}_{[b}\nabla^bU_{d]}=\epsilon
\mathcal{S}_d, \\
&&\label{E:ancorb} \nabla^a\left(\epsilon_{abcd}\mathcal{E}^c U^d+2 \mathcal{B}_{[a}U_{b]}\right)=\epsilon\mathcal{V}_b,
\eea
 where
\bea
\mathcal{S}_d\equiv2F^{e[b}R_{d]feb}-\epsilon\nabla_f J_d,
\quad
\mathcal{V}_b\equiv-\epsilon_f^{\phantom\ aud}F_{eu}R^e_{\phantom\ dab}.
\eea

\medskip
Projecting Eq.\il(\ref{E:treertb}) along  the  directions parallel and
orthogonal to the 4-velocity $U^b$, one obtains the following set of two
equations:
\bea
&& \label{E:UE1b}
\mathcal{E}_b\dot{U}^b+\epsilon \nabla^a\mathcal{E}_a-\epsilon_{abcd}U^a
\mathcal{B}^b\nabla^c U^d=-\epsilon U^b\mathcal{S}_b, \\\nonumber
&&
\dot{\mathcal{E}}_e+\epsilon U_e \mathcal{E}^b \dot{U}_b+\mathcal{E}_e \nabla^aU_a-h^b{}_e \mathcal{E}_a \nabla^aU_b-\epsilon U_e\epsilon_{abcd}U^b \mathcal{B}^c \nabla^a U^d\\\label{E:UE2b}
&&+\nabla^a(\epsilon_{aecd}\mathcal{B}^c U^d)=-\epsilon h^b{}_e \mathcal{S}_b.
\eea
A similar procedure applied to Eq.\il(\ref{E:ancorb}) gives
\bea
&& \label{E:Ub1b}
\epsilon_{abcd}U^b\mathcal{E}^c\nabla^a U^d+\epsilon
\nabla^a\mathcal{B}_a+B^b \dot{\mathcal{U}}_b=\epsilon
U^b\mathcal{V}_b, \\
\label{E:Ub2b}
&& \epsilon_{abcd}h^b{}_f\nabla^a(\mathcal{E}^cU^d)+\mathcal{B}_a h^b{}_f\nabla^a U_b-h^b{}_f \dot{\mathcal{B}}_b-\mathcal{B}_f \nabla^a U_a=\epsilon h_f{}^b\mathcal{V}_b.
\eea

\medskip
The propagation equations (\ref{E:UE1b})-(\ref{E:UE2b})  can be written  as
\bea\label{E:UE2ab}
\dot{\mathcal{E}}_{\langle f \rangle}&=&-2\mathcal{E}^a h_{f[a}\nabla_{b]} U^b-\epsilon_{abcd}h^b{}_f\nabla^a(\mathcal{B}^c U^d)+\epsilon h^d{}_f\mathcal{S}_d,
\\
\label{E:Ub2ab}
\dot{\mathcal{B}}_{\langle f \rangle}&=&-2\mathcal{B}^a h_{f[a}\nabla_{b]} U^b+\epsilon_{abcd}h^b{}_f\nabla^a(\mathcal{E}^cU^d)-\epsilon h^d{}_f\mathcal{V}_d.
\eea
while the constraint equations (\ref{E:Ub1b})-(\ref{E:Ub2b}) assume the form
\bea\label{E:Ub1ab}
\epsilon D^a\mathcal{B}_a&=&-\epsilon_{abcd}U^b\mathcal{E}^c\nabla^a U^d+\epsilon U^d\mathcal{V}_d,
\\
\label{E:UE1ab}
\epsilon D^a\mathcal{E}_a&=&\epsilon_{abcd}U^a \mathcal{B}^b\nabla^c U^d-\epsilon U^d\mathcal{S}_d.
\eea

\medskip
In order to verify that Eqs. (\ref{E:UE2ab})-(\ref{E:Ub2ab}) give
  appropriate hyperbolic evolution equations, we need the explicit
  expression for the electromagnetic current vector $J^a$ given
  by Ohm's law ---see Eq. (\ref{E:lg}). A computation then gives that
\bea
&& \mathcal{V}_bh^b{}_v =-\epsilon_f^{\phantom\ aud}F_{eu}R^e_{\phantom\
  dab}h^b_v. \label{B:2p2}\\\nonumber
&&\mathcal{S}_dh^d{}_v=F^{eb}R_{dfeb}h^d{}_v+R_{fe}\left(F^e_{\phantom\ v}-\epsilon E^e U_v\right)-\epsilon \left(
\nabla_f (\sigma E_v)+\epsilon U_v \sigma E_d
\nabla_fU^d+\rho_{\ti{C}} \nabla_fU_v\right). \label{E:2p2}
\eea
In the last equation it is assumed that the derivatives of the components of the Faraday tensor appearing in
the right hand side of (\ref{E:2p2}) are rewritten in terms of the
tensor $\psi_{abc}$. Furthermore, $R_{abcd}$ is expressed in terms of
its irreducible components.

\medskip
Finally, using  the kinematic quantities defined in
Eq.\il(\ref{E:dozero}) we finally rewrite Eqs.\il(\ref{E:UE2ab}-\ref{E:Ub2ab}) as:
\bea\label{E:UE2abz}\nonumber
\dot{\mathcal{E}}_{\langle f \rangle}&=&-\mathcal{E}^a \left[\chi h_{fa}-h_{fb}\left(\epsilon U_a a^b+\chi^b{}_a\right)\right]+\mbox{curl}\mathcal{B}_f-\epsilon\epsilon_{afcd}\mathcal{B}^c U^aa^d+\epsilon h^d{}_f\mathcal{S}_d,
\\\label{E:UE2abz}\\
\nonumber
\dot{\mathcal{B}}_{\langle f \rangle}&=&-2\mathcal{B}^a \left[\chi h_{fa}-h_{fb}\left(\epsilon U_a a^b+\chi^b{}_a\right)\right]-\mbox{curl}\mathcal{E}_f+\epsilon\epsilon_{afcd}\mathcal{E}^c U^aa^d-\epsilon h^d{}_f\mathcal{V}_d,
\\\label{E:Ub2abz}
\eea
where $ \mathcal{V}_bh^b{}_v$ and $\mathcal{S}_dh^d{}_v$ are given by
(\ref{B:2p2})-(\ref{E:2p2}). These equations provide the required
evolution equations for the components of the auxiliary field
$\psi_{abc}$. The hyperbolicity follows by comparison with the Maxwell
evolution equations, Eqs. (\ref{E:UE2a})-(\ref{E:Ub2a}).

\subsection{Propagation equations for frame coefficients and connection coefficients}

We now consider the evolution equations for the tetrad coefficients,
$e_a{}^\mu$, and the Ricci rotation coefficients,
$\Gamma_b{}^a{}_c$.

\medskip
The gauge choice $N=e_0=U$, $N_a=U_a$, $U^a=e_0{}^a=\delta_0{}^a$ together with  Eq.\il(\ref{E:h2o1}) readily give
\be\label{E:h2o1oboe}
\partial_t e_i{}^\mu=(\Gamma_{0\ i}^{\phantom\ j }-\Gamma_{i\ 0}^{\phantom\ j })e_j{}^\mu+\Gamma_{0\ i}^{\phantom\ 0 }e_0{}^\mu.
\ee

\medskip
In order to obtain evolution equations for the Ricci rotation
coefficients, one imposes the gauge condition $\Gamma_0{}^j{}_i=0$ arising from the Fermi propagation of the spatial
frame vectors ---see \cite{Fri98c}. It can be verified that
as a consequence of the metric compatibility condition
$\Gamma_{a(bc)}=0$ one only needs to consider evolution equations for
 the component $\Gamma_{q\ j}^{\phantom\ i }$, $\Gamma_{0\
i}^{\phantom\ 0 }$ and $\Gamma_{i\ j}^{\phantom\ 0 }$.

\medskip
A propagation equation for $\Gamma_{q\ j}^{\phantom\ i }$ is readily
found from Eq.\il(\ref{E:D41}). One has that
\be\label{E:rue1}
\partial_t \Gamma_{q\ j}^{\phantom\ i }=R^i_{j0q}-\Gamma_{k\ j}^{\phantom\ i }\Gamma_{q\ 0}^{\phantom\ k}-\Gamma_{0\ 0}^{\phantom\ i }\Gamma_{q\ j}^{\phantom\ 0}+\Gamma_{0\ j}^{\phantom\ 0 }\Gamma_{q\ 0}^{\phantom\ i }.
\ee

\medskip
The propagation equations for $\Gamma_{0\ i}^{\phantom\ 0 }$ and
$\Gamma_{q\ j}^{\phantom\ 0 }$ require a more elaborated analysis
involving the fluid equations. From Eq.\il(\ref{E:Tor}) together with the thermodynamic relations \il(\ref{E:the}) we obtain
\be\label{E:thez} \dot{p}=-\nu_s^2 \nabla^aU_a (\rho+p)-U^b
F_b^{\phantom\ c}\nabla^aF_{ac}.
\ee
Furthermore, using
Eq.\il(\ref{Eq:conservazione}) and Eq.\il(\ref{E:thez}) we have
\be
\label{E:Esta} \epsilon\nabla_dp=(\rho+p)(\dot{U}_d-U_d\nu_s^2
\nabla^a U_a)-\epsilon\nabla^aF_{ac}F_d^{\phantom\ c},
\ee
so that the condition $\nabla_{[a}\nabla_{b]}p=0$ with Eq.\il(\ref{E:Esta}) implies
\be\label{E:Estab}
0=\nabla_e\left((\rho+p)(\dot{U}_d-U_d\nu_s^2 \nabla^a U_a)\right)-\nabla_d\left((\rho+p)(\dot{U}_e-U_e\nu_s^2 \nabla^a U_a)\right)-2\epsilon\left(F_{[d}^{\phantom\ c}\nabla_{e]}+\nabla_{[e}F_{d]}^{\phantom\ c}\right)\nabla^aF_{ac}.
 \ee
This equation can be conveniently split as the sum of the three terms:
\be\label{E:Estaz}
\mathcal{J}_{ed}\equiv \mathcal{A}_{ed}+\mathcal{M}_{ed}+\mathcal{F}_{ed}=0,
\ee
where
\bea
 &&
\mathcal{A}_{ed}\equiv2(\rho+p)\nabla_{[e}(\dot{U}_{d]}-U_{d]}\nu_s^2
\nabla^a U_a),\\ && \mathcal{M}_{ed}\equiv 2(\dot{U}_{[d}-U_{[d}\nu_s^2
\nabla^a U_{|a|})\nabla_{e]}(\rho+p), \\
&&\mathcal{F}_{ed}\equiv-2\epsilon\left(F_{[d}^{\phantom\
c}\nabla_{e]}+\nabla_{[e}F_{d]}^{\phantom\ c}\right)\nabla^aF_{ac}.
\eea
Now, if one regards the velocity of sound $\nu_s$ as function of
$n$ and $s$, then from the $(0,i)$ component of Eq.\il(\ref{E:Estaz})
and Eq.\il(\ref{E:Esta}) we obtain:
\bea\nonumber
&&e_0 \Gamma_{0\ i}^{\phantom\ 0}-\nu^2_s e_i  \Gamma_{j\ 0}^{\phantom\ j}+
\Gamma_{0\ j}^{\phantom\ 0}\left( \Gamma_{i\ 0}^{\phantom\ j}-
\Gamma_{0\ i}^{\phantom\ j}\right)-\nu^2_s \Gamma_{j\ 0}^{\phantom\ j}
\Gamma_{0\ i}^{\phantom\ 0}=
\\\nonumber
&&-\frac{\rho+p}{\nu_s^2}\left(\frac{\partial^2p}{\partial \rho^2}\right)_s
\g{j}_{j\ 0}\left(\g{0}_{0\ i}+\frac{1}{p+\rho}(\nabla^{d}F_{de})F_{i}^{\phantom\ e}\right)
-\frac{\beta}{\rho+p}\g{0}_{0 \ i}\,\nabla_0 s
+\frac{\alpha}{\rho+p}\nabla_is\g{j}_{j\ 0}\\
&&-\frac{1}{\rho+p}(\g{0}_{0 \ i}\gamma_0-\g{j}_{j \ 0}\gamma_i)
-\frac{2}{\rho+p}\nabla_{[0}(F_{i]}^{\phantom\ c}\nabla^dF_{dc}),\label{E:Moz}
\eea
where
\bea && \label{E:ser} \gamma_a\equiv
\frac{\nu_s^2+1}{\nu_s^2}F_{a}^{\phantom\ c}\nabla^dF_{dc}, \\
&&\beta\equiv n T-\frac{1}{\nu_s^2}\frac{\partial p}{\partial
s},\qquad \alpha\equiv(\rho+p)\frac{\partial\nu_s^2} {\partial
s}-\left(1+\frac{n}{\nu_s^2}\frac{\partial\nu_s^2}{\partial
n}\right)\frac{\partial p}{\partial s}+\nu_s^2 n T.
\eea
Similarly,
from the $(j,k)$ component of Eq.\il(\ref{E:Estaz}) we find that

\bea\nonumber
&&e_k\Gamma_{0\ j}^{\phantom\ 0}-e_j\Gamma_{0\ k}^{\phantom\ 0}
-\Gamma_{0\ i}^{\phantom\ 0}\left(\Gamma_{k\ j}^{\phantom\ i}-\Gamma_{j\ k}^{\phantom\ i}\right)
-\nu_s^2 \Gamma_{i\ 0}^{\phantom\ i}\left(\Gamma_{k\ j}^{\phantom\ 0}-\Gamma_{j\ k}^{\phantom\ 0}\right)
\\
&&
+\frac{2\beta}{\rho+p}\nabla_{[k}s \, \Gamma_0{}^0{}_{j]}+\frac{2}{\rho+p}\g{0}_{0\ [k}\gamma_{j]}
+\frac{2}{\rho+p}\nabla_{[k}(F_{j]}^{\phantom\ c}\nabla^dF_{dc})=0. \label{E:Moz0}
\eea
Finally, making  use of Eq.\il(\ref{E:rue1}) in Eq.\il(\ref{E:Moz}) one obtains
\bea\nonumber
&&e_0\g{0}_{0\ i}-\nu_s^2 e_j\g{j}_{i\ 0}=
\g{0}_{0\ j}\left(\g{j}_{0\ i}-\g{j}_{i\ 0}\right)+\nu_s^2\g{j}_{j\  0}\g{0}_{0\ i}-\frac{\rho+p}{\nu_s^2}
\left(\frac{\partial^2p}{\partial \rho^2}\right)_s\g{j}_{j\ 0}
\\
\nonumber
&& \hspace{2cm}
+\left(\g{0}_{0\ i}+\frac{1}{p+\rho}(\nabla^{d}F_{de})F_{i}^{\phantom\ e}\right)
-\frac{\beta}{\rho+p}\g{0}_{0 \ i}\nabla_0 s
+\frac{\alpha}{\rho+p}\nabla_is\g{j}_{j\ 0}-\frac{1}{\rho+p}
(\g{0}_{0 \ i}\gamma_0-\g{j}_{j \ 0}\gamma_i)
\\
&& \hspace{2cm}-\frac{2}{\rho+p}\nabla_{[0}(F_{i]}^{\phantom\ c}\nabla^dF_{dc})+
\nu_s^2\left(R^{j}{}_{0ij}+\g{j}_{0 \ 0}(\g{0}_{i\ j}-\g{0}_{j\ i})-\g{j}_{p\ 0}\g{p}_{j \ i }
+\g{j}_{j \ p}\g{p}_{i \ 0}\right),
\eea
while using
\be\label{E:rue2}
\partial_t \Gamma_{k\ j}^{\phantom\ 0}=e_k \Gamma_{0\ j}^{\phantom\ 0}+R^{0}_{j0k}+ \Gamma_{0\ j}^{\phantom\ 0} \Gamma_{0\ k}^{\phantom\ 0}-\Gamma_{i\ j}^{\phantom\ 0}\Gamma_{k\ 0}^{\phantom\ i}-\Gamma_{0\ i}^{\phantom\ 0}\Gamma_{k\ j}^{\phantom\ i}
\ee
from Eq.\il(\ref{E:D41})
 in Eq.\il(\ref{E:Moz0}) gives
\bea\nonumber
&& e_0\g{0}_{k\ j}-e_{j}\g{0}_{0\ k}=
-\g{0}_{0\ p}(\g{p}_{j\ k}-\g{p}_{k\ j})
-\nu_s^2 \g{i}_{i\ 0}(\g{0}_{j\ k}-\g{0}_{k\ j})-\frac{2\beta}{\rho+p}\nabla_{[k}s\g{0}_{0\ j]}
\\
&&\nonumber
 -\frac{2}{\rho+p}\g{0}_{0\
[k}\gamma_{j]}-\frac{2}{\rho+p}\nabla_{[k}(F_{j]}^{\phantom\
c}\nabla^dF_{dc}) +R^{0}{}_{j0k}+\g{0}_{0\ j}\g{0}_{0\ k}-\g{0}_{i\
j}\g{i}_{k \ 0} -\g{0}_{0\ i}\g{i}_{k\ j}.
\\ \eea

In all the previous it is understood that terms of the form $\nabla^dF_{dc}$ are to be replaced using
the relation

\be
\nabla^dF_{dc}=\psi_c=\rho_{\ti{C}}U_c+\sigma E_c,\label{E:sort}
\ee
where it is assumed that the charge density, $\rho_{\ti{C}}$, is
constant multiple of the fluid  density, $\rho=\varrho
\rho_{\ti{C}}$.

\medskip
\noindent
\textbf{Remark.} Important for the hyperbolicity of the evolution
equations is that the term
\[
\frac{2}{\rho+p}\nabla_{[a}(F_{b]}^{\phantom\ c}\nabla^dF_{dc}),
\]
appearing in Eqs.\il(\ref{E:Moz}) and Eq.\il(\ref{E:Moz0}), can be
rewritten in terms of $(\g{a}_{c\ d},s_a,\mathcal{E}_a)$ if one uses
Eqs.\il(\ref{PsiElectric-Magnetic}) and \il(\ref{E:sort}).

\subsection{Summary of the analysis}
\label{Section:Summary}

In this subsection we summarise our analysis of the evolution
equations. Evolution equations for the \emph{independent} components of the vector
variable
\be\label{E:v} \vec{v}=\left(e_i{}^\mu,\g{0}_{0\ i},\g{a}_{i\
b},\hat{E}_{ab},\hat{B}_{ab},E_{a},B_{a},\mathcal{E}_{a},
\mathcal{B}_{a}, n, s , s_a\right),
\ee
have been constructed. The components of $\Gamma_i{}^j{}_k$ not
included in this list are determined by means of gauge conditions and
symmetries. By construction, the electric and magnetic parts of the
Weyl tensor are tracefree. This symmetry is disregarded in these
considerations and is recovered by imposing it on the initial data. It
can be shown that if these tensors are initially tracefree, then they
will be also tracefree for all later times ---see
e.g. \cite{FriRen00}.

\medskip
The evolution equations for the independent components of the unknowns
in (\ref{E:v}) and the underlying assumptions are given as follows:

\begin{itemize}
\item[(i)] The propagation equation for the tetrad coefficients,
$e_i{}^\mu$, is given by Eq.\il(\ref{E:h2o1oboe}) by virtue of the
Lagrange condition $U^a=e^a_0=\delta^a_0$. Eq.\il(\ref{E:h2o1oboe})
has the same principal part than the corresponding equation in the
uncharged case as described in \cite{Fri98c}. It gives rise
to a symmetric hyperbolic subsystem of equations.

\item[(ii)] The connection coefficients, $\g{0}_{0\ i}$ and $\g{a}_{i\
b}$ are given, respectively, by Eqs. \il(\ref{E:Moz}) and
Eq.\il(\ref{E:Moz0}). In addition, the gauge condition, $\Gamma_{0\
i}^{\phantom\ j }=0$, is assumed.  Eq.\il(\ref{E:rue1}) takes care of
$\g{q}_{j\ k}$. As in the case of the equations frame coefficients,
Eqs. \il(\ref{E:Moz}), \il(\ref{E:Moz0}) and \il(\ref{E:rue1}) have
the same principal part as in the uncharged case analysed in
\cite{Fri98c}. These equations again form a symmetric
hyperbolic subsystem of equations.

\item[(iii)] The evolution equations for the electric part,
$\hat{E}_{ab}$, and the magnetic part, $\hat{B}_{ab}$, of the Weyl
tensor are given, respectively, by Eqs. (\ref{E:terete}) and
\il(\ref{E:sa}). As mentioned before, the tracefreeness of these
tensors is not used to reduce the number of independent
components. Thus one has 12 equations for equal number of
components. Equations with a principal part of the form of
Eqs. (\ref{E:terete}) and \il(\ref{E:sa}) are symmetric hyperbolic
independently of the gauge choice ---see e.g. \cite{FriRen00}.

\item[(iv)] The propagation equation for the electric part, $E_a$,
and the magnetic part, $B_b$, of the Faraday tensor are given,
respectively, by Eqs.\il(\ref{E:UE2a}) and \il(\ref{E:Ub2a}). The isotropic Ohm
law, Eq. (\ref{E:lg}) has been assume in the deduction of these equations. As in the
case of the equations for the electric and magnetic parts of the Weyl
tensor, the principal part of these equations is known to be
hyperbolic independently of the gauge ---again, see
e.g. \cite{FriRen00}.

\item[(v)] The evolution equations for the electric part,
$\mathcal{E}_a$, and magnetic part, $\mathcal{B}_b$, of the field
$\psi_{abc}$, corresponding to the covariant derivative of the Faraday
tensor are given, respectively, by Eqs.\il(\ref{E:UE2abz}) and
\il(\ref{E:Ub2abz}). These equations involve 24 equations for as many
unknowns. Their structure is analogous to that of
Eqs. \il(\ref{E:UE2a}) and \il(\ref{E:Ub2a}), except that they
contain an extra free index. As a consequence, their principal part
gives rise to a symmetric hyperbolic subsystem.

\item[(vi)] The evolution equation for the particle number density $n$ is
given by Eq.\il(\ref{E:then}). This equation is consistent with
symmetric hyperbolicity as in the Lagrangian gauge all the derivatives
of the flow vector $U^a$ are replaced by connection coefficients, and
thus, it contains no derivatives in the spatial directions.

\item[(vii)] The evolution equation for the entropy per particle, $s$, is
given by Eq.\il(\ref{E:fors}) with the understanding that the
covariant derivative of the Faraday tensor arising in the right-hand
side is to be expressed in terms, either, of the auxiliary field
$\psi_{abc}$ or written in terms of the current vector $J_a$ using the
inhomogeneous Maxwell equation (\ref{E:MW}).

\item[(viii)] Finally the equation for $s_a\equiv\nabla_a s$ is
obtained by covariant differentiating Eq.\il(\ref{E:ser}), commuting
covariant derivatives and taking into account having in mind
Eqs.\il(\ref{E:sort}) and, again, \il(\ref{E:ser}).

\end{itemize}

\medskip
Summarising,  Eqs. (\ref{E:UE2a}), (\ref{E:Ub2a}), (\ref{E:then}), (\ref{E:sa}), (\ref{E:terete}),
(\ref{E:UE2abz}), (\ref{E:Ub2abz}), (\ref{E:h2o1oboe}),
(\ref{E:rue1}), (\ref{E:Moz}), (\ref{E:Moz0}) and (\ref{E:ser})
provide the desired symmetric hyperbolic evolution system for the
Einstein-charged perfect fluid system. This system can be written in
the form
\begin{equation}
\label{E:II}
\mathbf{A}^0\partial_0 \vec{v}-\mathbf{A}^j\partial_j \vec{v}=\mathbf{B} \vec{v},
\end{equation}
where $\mathbf{A}^\mu=\mathbf{A}^0(x^{\mu},\vec{v})$ are matrix-valued
function of the coordinates and the unknowns $\vec{v}$, and
$\mathbf{A}^{0}$ is positive definite. The structure of the system
(\ref{E:II}) ensures the existence of local solutions to the evolution
equations. The analysis of whether these solutions give rise to a
solution of the full Einstein-charged perfect fluid system requires
the analysis of the propagation of the constraint equations. This is a
computationally intensive argument which will be omitted here.

\subsection{Infinitely conductive plasma}
\label{Subsection:MHD}

A particularly important subcase of our previous analysis is that of
an ideal conductive plasma, defined by the condition  $E_a=0$ ---ideal
magnetohydrodynamics. The
hyperbolic problem for this case can be naturally treated as a subcase
of the general case in which the Faraday tensor has all  the six
independent components. In this case the vector variable $\vec{v}$
reduces to
\be\label{E:vE0}
\vec{v}=\{e^{\mu}_i,\g{0}_{0\ i},\g{a}_{i\ b},\hat{E}_{ab},\hat{B}_{ab},B_{a},\mathcal{E}_{a},
\mathcal{B}_{a}, n, s , s_a\}.
\ee
Notice that although $E_a=0$, one nevertheless has a non-vanishing
electric part of the auxiliary tensor $\psi_{abc}$ as can be seen from $\mathcal{E}_i=-\g{j}_{a\ d}F_{ij}U^d$.

\medskip
In the infinitely conductive plasma case  the system consisting
of Eq.\il(\ref{E:MW}), Eq.\il(\ref{Eq:conservazione}),
Eq.\il(\ref{E:qwety}), and Eq.\il(\ref{E:fors}) reduces to
\begin{eqnarray}
&& U_a\nabla^a\rho+(p+\rho)\nabla^aU_a=0, \label{E:1}
\\
&& (p+\rho)U^a\nabla_aU^c-\epsilon h^{bc}\nabla_b p-\epsilon(\nabla^aF_{ad})F^{\phantom\ d}_b h^{bc}=0,
\\
&& U^a\nabla_a s=0, \label{E:somo}
\end{eqnarray}

\medskip
\noindent
\textbf{Remark 1.} As a consequence of Eq.\il(\ref{E:somo}) the
entropy per particle is constant along the flow lines of the perfect
fluid. This fact is translated into Eq.\il(\ref{E:somo}), propagation
equation for $s$, that implies the propagation equation $\mathcal{L}_U
s_a=0$.

\medskip
\noindent
\textbf{Remark 2.} Consistent with Ohm's law the source term of the
Maxwell equation is in this case simply given by $J_a=\rho_{\ti{C}}
U_a$.

\section{Conclusions}\label{Sec:Colcusione}

In the present article we have revisited the issue of wellposedness
initial value problem for the evolution equations of the
Einstein-Maxwell-Euler system (a selfgravitating charged perfect
fluid). Our analysis assumes a system consisting of a single
relativistic perfect fluid obeying to a certain equation of state. The
approach followed makes use of the well known $1+3$ tetrad formalism
by means of which the various tensorial quantities and equations are
projected along the direction of the comoving observer and onto the
orthogonal subspace. Following \cite{Fri98c,FriRen00}, we require the
timelike vector of the orthonormal frame to follow the matter flow
lines (Lagrangian gauge).  Moreover, we assume the vector fields
tetrad to be Fermi transported in the direction of $U$, these
conditions fix certain components of the connection.

\medskip
 A key feature of our analysis was the introduction
of a  rank 3 tensor $\psi_{abc}$ corresponding to the covariant
derivative of the Faraday tensor $F_{ab}$. The purpose of introducing
this tensor was to ensure the symmetric hyperbolicity of the
propagation equations for the components of the Weyl tensor. The
symmetries of the tensor $\psi_{abc}$ readily allow a decomposition
analogous to that of the Faraday tensor into electric and magnetic
parts.

\section*{Acknowledgments}
DP gratefully acknowledges financial support from the A. Della Riccia
Foundation. Part of this project was carried out while JAVK attended
the programme on ``The Dynamics of the Einstein Equations'' at the
International Erwin Schro\"odinger Institute in Vienna, Austria from
July to September 2011. The hospitality of the Institute is kindly acknowledged.

\appendix

\section{On the equations of state}\label{E:Aa}

\subsection{Homentropic flows}

A fluid is said to be \emph{homeontropic} is $s$ is constant in space
and time. In general, the equation of state is given by $\rho=
f(n,s)$. Now, if one has an homeontropic flow, the latter can be
written as $\rho=f(n)$ ---there is no dependence on $s$ as it is
constant. Now, if $f$ is a differentiable function of $n$ and
$f'(n)\neq 0$, then one can write $n =f^{-1}(\rho)$. It then follows
that
\begin{eqnarray*}
&& p = n \frac{\partial \rho}{\partial n} - \rho  , \\
&& \phantom{p} = n f'(n) - \rho, \\
&& \phantom{p}= n f'(f^{-1}(\rho))-\rho.
\end{eqnarray*}
Thus, $p$ is of the form $p=h(\rho)$ ---that is, one obtains a
barotropic equation of state.

\subsection{Conditions for a barotropic equation of state}
In the previous section we have shown that the assumption of an
homeontropic flow implies a barotropic equation of state. We now
investigate more general situations for which one can have this type
of equation of state.

Assume that $p=h(p)$ . Then, from
\[
p(n,s) = n \left( \frac{\partial \rho }{\partial n}\right) - \rho(n,s)
\]
it follows that
\[
h( f(n,s)) = n \left( \frac{\partial f }{\partial n}\right) - f(n,s).
\]
The latter can be read as a (possibly non-linear) differential
equation  for $f(n,s)$. It can be integrated to obtain
\[
\exp\left( \int \frac{\mbox{df}}{h(f)+f}  \right) = g(s) n,
\]
where $g(s)$ is an arbitrary function of $s$. This expression can be
used to eliminate $n$ from the discussion.  In the case of a dust
equation of state ($p=0$), the last relation reduces to
\[
\rho/ n = g(s).
\]

\addcontentsline{toc}{chapter}{Bibliography}

\end{document}